\documentclass[review]{elsarticle}

\usepackage{lineno}
\usepackage{booktabs}
\usepackage{lineno}
\usepackage{graphicx}
\usepackage{multirow}
\usepackage[dvipsnames]{xcolor}
\usepackage{balance}
\usepackage{bm}
\usepackage[hyphens]{url}
\usepackage[font=footnotesize,labelfont=bf]{caption}
\usepackage{array}
\usepackage{amsmath}
\modulolinenumbers[5]

\journal{Journal of Information Processing and Management}

\biboptions{authoryear}

\begin{document}

\begin{frontmatter}

\title{User and Item-aware Estimation of Review Helpfulness}


\author[mymainaddress]{Noemi Mauro}
\ead{noemi.mauro@unito.it}

\author[mymainaddress]{Liliana Ardissono\corref{mycorrespondingauthor}}
\ead{liliana.ardissono@unito.it}

\author[mymainaddress]{Giovanna  Petrone}
\ead{giovanna.petrone@unito.it}

\cortext[mycorrespondingauthor]{This is to indicate the corresponding author.}

\address[mymainaddress]{Dipartimento di Informatica, Universit\`{a} degli Studi di Torino, Corso Svizzera 185, I-10149 Torino, Italy}

\begin{abstract}
In online review sites, the analysis of user feedback for assessing its helpfulness for decision-making is usually carried out by locally studying the properties of individual reviews.
However, global properties should be considered as well to precisely evaluate the quality of user feedback. 

In this paper we investigate the role of {\em deviations} in the properties of reviews as helpfulness determinants with the intuition that ``out of the core'' feedback helps item evaluation. We propose a novel helpfulness estimation model that extends previous ones with the analysis of {\em deviations in rating, length and polarity} with respect to the reviews written by the same person, or concerning the same item. A regression analysis carried out on two large datasets of reviews extracted from Yelp social network shows that user-based deviations in review length and rating clearly influence perceived helpfulness.
Moreover, an experiment on the same datasets shows that the integration of our helpfulness estimation model improves the performance of a collaborative recommender system by enhancing the selection of high-quality data for rating estimation. Our model is thus an effective tool to select relevant user feedback for decision-making.
\end{abstract}


\end{frontmatter}
Declarations of interest: none.
\\
\\
\textbf{Published in Information Processing \& Management, Elsevier. 
\\
DOI: \url{https://doi.org/10.1016/j.ipm.2020.102434}. 
\\Link to the page of the paper on Elsevier web site:
\\
\url{https://www.sciencedirect.com/science/article/pii/S0306457320309274}
\\
This work is licensed under the:
\\Creative Commons Attribution-NonCommercialNoDerivatives 4.0 International License. 
\\
To view a copy of this license, visit  \url{http://creativecommons.org/licenses/by-nc-nd/4.0/} or send a letter to Creative Commons, PO Box 1866, Mountain View, CA 94042, USA. } 


\section{Introduction}
\label{sec:introduction}
Reviews of items posted in e-commerce sites and social media are a precious source of information about consumers' experience with products but their abundance challenges their effective fruition.
Marketers thus attempt to evaluate the helpfulness of reviews in order to promote those which best support purchasing decisions.
However, depending on factors such as age (recentness) and level of visibility in the web sites, good reviews might fail to get feedback from readers \citep{Hu-Chen:16,Hu-etal:17}. Therefore, helpfulness must be estimated {\em a priori} to sort comments in an informed way as soon as they are posted \citep{Krestel-Dokoohaki:15}. 

Several researchers assume that helpfulness is an internal property of reviews. For instance, see \cite{Mudambi-Schuff:10}, \cite{Yang-etal:15}, \cite{Hong-etal:17} and \cite{Siering-etal:18}. In these works, each comment is analyzed independently of the others. However, linguistic style is personal \citep{Li-etal:19} and perceived helpfulness also depends on the variability of ratings provided by reviewers \citep{Gao-etal:17}.
Moreover, \cite{Raghavan-etal:12} and \cite{Fang-etal:16} observed that the deviation with respect to the mean rating of a product supports helpfulness estimation. 
Starting from these findings, we are interested in understanding whether a contextual analysis of reviews written by the same person, or concerning the same item, contributes to enhance helpfulness assessment. 
Specifically, we investigate the role of deviations in the content properties of reviews as helpfulness determinants. We pose the following research questions:
\begin{itemize}
    \item[RQ1:]
    {\em Given a review r, does a deviation from the mean length, polarity and rating of the other reviews written by the same person provide useful information to assess the perceived helpfulness of r?}
    \item[RQ2:] 
    {\em Given a review r, does a deviation from the mean length, polarity and rating of the other reviews on the same item provide useful information to assess the perceived helpfulness of r?}
\end{itemize}
In order to answer these questions we analyzed two datasets of reviews from \cite{Yelp}, one about accommodation services (including about 10000 comments) and the other about food services (including about 65000 comments). We learned through regression a helpfulness estimation model that combines (i) largely used determinants such as review length, TF/IDF statistics and ratings with (ii) review polarity, that is the sentiment emerging from its text, and (iii) the deviations of these features among the comments provided by the same user, or concerning a single item. Then, we compared the helpfulness estimation capability of our model with that of baseline models that only use factors of type (i) and (ii).
We carried out the evaluation as follows:
\begin{enumerate}
    \item 
    First, we checked whether our model estimates helpfulness more accurately than the baselines by correlating the predicted values with the feedback about reviews observed in the datasets (ground-truth helpfulness) by means of Pearson and Spearman analyses.
    \item 
    Then, we evaluated whether our model supports the identification of high-quality information for decision-making. We did this by extending a collaborative recommender system \citep{Koren-Bell:11} to weight the impact of observed rating data on recommendation, and by comparing suggestion performance with that of standard Collaborative Filtering. 
\end{enumerate}
On both datasets the experimental results show that our model better adheres to ground-truth helpfulness than the baselines. Moreover, our model enhances recommendation performance in terms of accuracy, error minimization and ranking of items. 
In summary, we provide the following contributions:
\begin{itemize}
    \item 
    An advancement of the state of the art in review helpfulness estimation based on the idea that ``out of the core'' reviews can be relevant information sources for decision-making.
    \item
    A novel helpfulness prediction model that extends previous ones with the identification of user-based and item-based helpfulness determinants.
    \item 
    An experimental validation which shows that our model outperforms the selected baselines and improves collaborative item recommendation.
\end{itemize}
Our work paves the way toward the development of human-centered algorithms by enhancing performance and transparency of recommender systems. Specifically, review helpfulness prediction can be used to select high-quality ratings for recommendation. Moreover, it can be employed to explain the suggestions generated by the system using the textual feedback provided by previous consumers to describe their experience with items \citep{Ghose-Ipeirotis:11}.

The remainder of this article is organized as follows. Section \ref{sec:related} provides a literature review. Section \ref{sec:method} presents our research methodology and Section \ref{sec:results} describes the experimental results. Section \ref{sec:recommender} shows the benefits of our model to personalized item recommendation. Section \ref{sec:discussion} summarizes the findings of our study and their implications. Section \ref{sec:future} describes limitations and suggestions for future work and Section \ref{sec:conclusions} concludes the paper.

\section{Background and Related Work}
\label{sec:related}
This section positions our work with respect to the research about helpfulness determinants concerning review content. See \cite{Ocampo-Diaz-Ng:18} for a survey. Table \ref{tab:research} summarizes these factors by reporting, for each one, the impact on perceived helpfulness and the works supporting the finding. 
Most of the cited works study a larger set of determinants but Table \ref{tab:research} only shows those identified as influential by the authors of the cited works.

\begin{table*}[!t]
\centering 
\caption{Determinants of perceived review helpfulness concerning review content. }
\resizebox{\columnwidth}{!}{%
{\def\arraystretch{2.4}
\begin{tabular}{llllll}
\toprule
\vspace{4mm}
Determinant & Definition & Prior finding & Represented study

\\
\midrule

length (depth) & 
 \multicolumn{1}{l}{
 \renewcommand{\arraystretch}{0.5}
 \begin{tabular}[l]{@{}l@{}}
 total number of \\
 words in a review
 \end{tabular}}
 &
 \multicolumn{1}{l}{
 \renewcommand{\arraystretch}{0.5}
 \begin{tabular}[l]{@{}l@{}} \hspace{-2.3mm}
 positive, negative,\\ \hspace{-2.3mm}
 inverted-U-shaped
 \end{tabular}} &
 \multicolumn{1}{l}{
 \renewcommand{\arraystretch}{0.5}
 \begin{tabular}[l]{@{}l@{}}
 \cite{Kim-etal:06},  \cite{Wu:17}, \\
 \cite{Mudambi-Schuff:10},\\
 \cite{Fink-etal:18}, \cite{Eslami-etal:18}\\
 \cite{Hong-etal:17}\\
 \cite{Raghavan-etal:12}
 \end{tabular}}\\
 
 Unigram & 
 \multicolumn{1}{l}{
 \renewcommand{\arraystretch}{0.5}
 \begin{tabular}[l]{@{}l@{}}
 TF/IDF value of\\
 the words included \\
 in the review
 \end{tabular}}
 & not specified &  
 \multicolumn{1}{l}{
 \renewcommand{\arraystretch}{0.5}
 \begin{tabular}[l]{@{}l@{}} 
 \cite{Kim-etal:06}, \cite{O'Mahony-Smyth:18}
 \end{tabular}}\\

entropy & 
\multicolumn{1}{l}{
\renewcommand{\arraystretch}{0.5}
 \begin{tabular}[l]{@{}l@{}}
 entropy of the words\\
 included in the review
 \end{tabular}}
& negative &  \hspace{0mm} \cite{Fresneda-Gefen:19}\\

rating & \hspace{0mm} star rating in [1, 5] & 
 \multicolumn{1}{l}{
 \renewcommand{\arraystretch}{0.5}
 \begin{tabular}[l]{@{}l@{}}
 \hspace{-2.3mm}
 positive,\\ \hspace{-2.3mm}
 inverted-U-shaped
 \end{tabular}} &
\multicolumn{1}{l}{
 \renewcommand{\arraystretch}{0.5}
 \begin{tabular}[l]{@{}l@{}}
 \cite{Kim-etal:06}, \cite{Eslami-etal:18}, \\
 \cite{Mudambi-Schuff:10}, \\
 \cite{O'Mahony-Smyth:10,O'Mahony-Smyth:18}
 \end{tabular}}\\ 
 
writing style & 
 \multicolumn{1}{l}{
 \renewcommand{\arraystretch}{0.5}
 \begin{tabular}[l]{@{}l@{}}
 readability, \\
 linguistic correctness
 \end{tabular}}
 &  
  \multicolumn{1}{l}{
 \renewcommand{\arraystretch}{0.5}
 \begin{tabular}[l]{@{}l@{}} \hspace{-2.3mm}
 positive, \\ \hspace{-2.3mm}
 domain dependent, \\ \hspace{-2.3mm}
 insignificant
 \end{tabular}}
 & 
 \multicolumn{1}{l}{
 \renewcommand{\arraystretch}{0.5}
 \begin{tabular}[l]{@{}l@{}}
 \cite{Ghose-Ipeirotis:11}, \cite{Liu-etal:18}\\
 \cite{Hong-etal:17}, \cite{Krishnamoorthy:15}
 \end{tabular}}\\

subjectivity & 
 \multicolumn{1}{l}{
 \renewcommand{\arraystretch}{0.5}
 \begin{tabular}[l]{@{}l@{}}
  subjective statements  \\
  in review 
 \end{tabular}}
& mix is negative &
 \multicolumn{1}{l}{
 \renewcommand{\arraystretch}{0.5}
 \begin{tabular}[l]{@{}l@{}}
  \cite{Ghose-Ipeirotis:11},  \\
  \cite{Krishnamoorthy:15}
 \end{tabular}} \\

linguistic features & 
 \multicolumn{1}{l}{
 \renewcommand{\arraystretch}{0.5}
 \begin{tabular}[l]{@{}l@{}}
  adjectives; state  \\
  and action verbs; \dots
 \end{tabular}}
& positive & \hspace{0mm} \cite{Krishnamoorthy:15}\\

semantic features & 
 \multicolumn{1}{l}{
 \renewcommand{\arraystretch}{0.5}
 \begin{tabular}[l]{@{}l@{}}
  total number of \\
  concepts in the review; \\
  average number of \\
  concepts per sentence
 \end{tabular}} &
 positive & 
  \multicolumn{1}{l}{
 \renewcommand{\arraystretch}{0.5}
 \begin{tabular}[l]{@{}l@{}}
  \cite{Qazi-etal:16}, \cite{Cao-etal:11b} \\
  \cite{Sun-etal:19}
 \end{tabular}} \\
 \multicolumn{1}{l}{
 \renewcommand{\arraystretch}{0.5}
 \begin{tabular}[l]{@{}l@{}} \hspace{-2.3mm}
  polarity (valence, \\ 
   \hspace{-2.3mm}
  sentiment)
 \end{tabular}} & 
 \multicolumn{1}{l}{
 \renewcommand{\arraystretch}{0.5}
 \begin{tabular}[l]{@{}l@{}}
  positive/negative \\
  sentiment of review
 \end{tabular}}
& negative &
 \multicolumn{1}{l}{
 \renewcommand{\arraystretch}{0.5}
 \begin{tabular}[l]{@{}l@{}}
  \cite{Eslami-etal:18}, 
  \cite{Dong-etal:13}, \\
  \cite{Siering-etal:18},
  \cite{Salehan-Kim:16}
 \end{tabular}}\\
 
aspects &  \multicolumn{1}{l}{
 \renewcommand{\arraystretch}{0.5}
 \begin{tabular}[l]{@{}l@{}}
  semantic features \\
  occurring in reviews
 \end{tabular}} 
 & depends on polarity &
 \multicolumn{1}{l}{
 \renewcommand{\arraystretch}{0.5}
 \begin{tabular}[l]{@{}l@{}}
  \cite{Paul-etal:17}, \cite{Xiong-Litman:14}, \\
  \cite{Yang-etal:15,Yang-etal:16}
 \end{tabular}} \\
 
coherence (consistency) & 
 \multicolumn{1}{l}{
 \renewcommand{\arraystretch}{0.5}
 \begin{tabular}[l]{@{}l@{}}
  consistency between \\
  review polarity and rating,\\
  similarity between review\\
  title and content
 \end{tabular}} 
 & --, negative & 
 \multicolumn{1}{l}{
 \renewcommand{\arraystretch}{0.5}
 \begin{tabular}[l]{@{}l@{}}
  \cite{Dong-etal:13}, \cite{Zhou-etal:20},\\
  \cite{Shen-etal:2019}
 \end{tabular}}  \\
 
rating deviation &
 \multicolumn{1}{l}{
 \renewcommand{\arraystretch}{0.5}
 \begin{tabular}[l]{@{}l@{}}
  deviation of rating from \\
  mean product rating
 \end{tabular}} &
 positive & \hspace{0mm} \cite{Raghavan-etal:12} \\

\multicolumn{1}{l}{
 \renewcommand{\arraystretch}{0.5}
 \begin{tabular}[l]{@{}l@{}}
 \hspace{-2.3mm}
 domain-specific \\
 \hspace{-2.3mm}
 content
 \end{tabular}} &
\multicolumn{1}{l}{
 \renewcommand{\arraystretch}{0.5}
 \begin{tabular}[l]{@{}l@{}}
 domain-specific\\
 item properties 
 \end{tabular}} & 
 \multicolumn{1}{l}{
 \renewcommand{\arraystretch}{0.5}
 \begin{tabular}[l]{@{}l@{}} \hspace{-2.3mm}
 moderated \\ \hspace{-2.3mm}
 by product 
 \end{tabular}} & \hspace{0mm}
 \cite{Ahmad-Laroche:17}\\
 
 age & 
  \multicolumn{1}{l}{
 \renewcommand{\arraystretch}{0.5}
 \begin{tabular}[l]{@{}l@{}}
  number of days \\
  since publication
 \end{tabular}}
 & positive & 
  \multicolumn{1}{l}{
 \renewcommand{\arraystretch}{0.5}
 \begin{tabular}[l]{@{}l@{}}
  \cite{Hong-etal:17}, \hspace{0mm} \citep{Hu-Chen:16} \\
 \end{tabular}}\\

\bottomrule
\end{tabular}
} }
\label{tab:research}
\end{table*}

\subsection{Review helpfulness for item recommendation}
Some researchers use reviews to calculate the associated ratings by means of a content analysis. For instance, see \cite{Margaris-etal:20}. Our work is different because we measure review helpfulness to {\em select} high-quality ratings for recommendation.
This selection is a pre-requisite for the generation of relevant suggestions because algorithmic accuracy is not sufficient when working with poor data. Moreover, this selection supports recommender systems transparency by identifying appropriate feedback about items that can be used to explain the generated results. This improves traceability and trust which, as discussed by \cite{Shin:20b}, enhance user acceptance of services. Specifically, \cite{Shin-etal:20} have discovered that algorithmic experience ``is inherently related to human understanding of fairness, transparency, and other conventional components of user-experience'' which, in turn, are tightly connected to explainability.

Recommender systems \citep{Ricci-etal:11} address transparency \citep{Tintarev-Masthoff:15} and trust \citep{Berkovsky-etal:17,Berkovsky-etal:18} by enriching the suggestions they generate with a description of the degree, or of the type of matching between users and items. For instance, see \cite{Herlocker-etal:00}, \cite{Kouki-etal:19} and \cite{Pu-Chen:07}. We claim that item reviews perceived as helpful by their readers are an important asset to be used for this purpose because they make it possible to describe item properties by exploiting previous consumer experience \citep{Mauro-etal:20d,Ghose-Ipeirotis:11}. This is in line with findings related to the news recommendation domain, in which \cite{Shin:20} has pursued interactivity and presented ``a news recommendation experience model incorporating algorithm quality (transparency and accuracy) and perceived value (utility and convenience) as antecedent factors of conﬁrmation and satisfaction.''

\subsection{Review-related helpfulness determinants}
\subsubsection{Structural features}
Review length, rating (number of stars) and Unigram (TF/IDF statistics of the words appearing in the review \citep{Robertson:04}) are recognized as important helpfulness determinants as found by \cite{Kim-etal:06}. Length is taken as a proxy of informativeness and can be associated to user involvement in writing the comment \citep{Pan-Zhang:11}. \cite{Fink-etal:18} observed that, when content is created under low-constraint settings, such as mobile interaction, length has an inverted-U-shaped influence on perceived helpfulness, with medium length comments being more effective than very short and very long ones. 
Rating is taken as a proxy of review valence representing positive/negative opinion. Unigram assesses the relevance of review words when compared to the other comments about the same product. It can be noticed that Unigram is not the only way to measure relevance. For example, \cite{Fresneda-Gefen:19} evaluated words ``unicity'' in terms of message entropy.

\cite{Ghose-Ipeirotis:11} analyzed the {\bf readability} of reviews and their {\bf linguistic correctness} (lack of misspellings, etc.), both of which are observed to positively influence perceived helpfulness. However, \cite{Liu-etal:18} proved that readability depends on how closely a review matches the language style of the target readers. Therefore it is a domain-dependent indicator. 

In order to base our work on largely applicable helpfulness determinants, we focus our analysis on length, rating and Unigram, leaving linguistic correctness and readability apart. Moreover, we study deviations in length and rating by grouping reviews by author or item in order to provide a contextual analysis of user feedback. Finally, we test the usefulness of our helpfulness estimation model for decision-making by integrating it into a collaborative recommender system and by measuring the improvements in suggestion performance.

\subsubsection{Semantic features}
\cite{Cao-etal:11b} and \cite{Qazi-etal:16} found that {\bf semantic features} of reviews positively influence helpfulness perception. Indeed, the semantic analysis includes diverse approaches that also exploit some structural features, such as the number of product attributes mentioned in a review, and the length of its sentences \citep{Sun-etal:19}. 
 
Among the identified helpfulness predictors there are the {\bf positive or negative sentiment (polarity)} of reviews, combined with the number of positive/negative words \citep{Dong-etal:13}. \cite{Eslami-etal:18} observed that the most helpful comments are associated to medium length, lower scores, and negative or neutral polarity. \cite{Ahmad-Laroche:17} noticed that negative reviews containing service failure data and positive reviews describing core product functionalities, technical aspects and aesthetics are perceived as helpful. \cite{Salehan-Kim:16} found that sentimental reviews with neutral polarity in their text are perceived to be more helpful than the other ones. 

Differently, \cite{Ghose-Ipeirotis:11} discovered that very {\bf objective} and very {\bf subjective} comments are considered as helpful but mixed comments are not. Moreover, \cite{Krishnamoorthy:15} observed that syntactic structure and presence of adjectives, state and action verbs are good helpfulness predictors, especially if used in conjunction with readability and subjectivity, review age and rating. 

{\bf Aspect-based approaches} for helpfulness assessment employ techniques such as Supervised LDA \citep{Blei-etal:10} and double propagation to extract aspects from reviews as latent topics. See \cite{Xiong-Litman:14} and \cite{Paul-etal:17}, respectively. However, \cite{Yang-etal:16} noticed that LDA produces a large number of low-level, product-dependent aspects. 

We aim at developing a model that can be transferred to different service domains. For this purpose, we focus on review polarity, which we analyze both in absolute terms, as done in previous work, and contextually, from the viewpoint of user/item-based deviations. Moreover, we integrate our model into collaborative recommendation.

\subsubsection{Consistency and Rating Deviations}
Some recommender systems use {\bf consistency} (henceforth, {\bf coherence}) to evaluate reviewers' reliability, having observed that large discrepancies between review sentiment and rating can be a sign of low-quality \citep{Shen-etal:2019}. Also \citep{Dong-etal:13} investigated this feature but they have not described its impact on perceived helpfulness.
\cite{Zhou-etal:20} have looked at consistency from a different perspective and they discovered that the similarity between review title and review content positively influences perceived helpfulness.

\cite{Raghavan-etal:12} found that review length and the {\bf deviation of the rating} from the mean rating of the product are strong helpfulness predictors. Moreover, they observed that the regression models that use these features perform better than those relying on semantic features, either based on TF/IDF or LDA.

In our helpfulness estimation model we include the rating-polarity coherence as a candidate determinant but we omit the analysis of the similarity between title and content because, as described in Section \ref{sec:data-collection}, the reviews used for our experiments have no title. However, our model could be seamlessly extended to consider this additional element.

\subsection{Moderating factors}
\label{sec:moderation}
Two {\em moderating factors} of helpfulness perception can influence readers' voting behavior:
\begin{itemize}
    \item 
   The first is {\bf Product type}. \cite{Mudambi-Schuff:10} observed that extreme ratings negatively influence perceived helpfulness in {\em experience goods}, while review length has greater positive effect on {\em search goods} than on experience ones. Moreover, \cite{Siering-etal:18} found that the strength of sentiment increases review helpfulness for search products while it decreases helpfulness for experience products.\footnote{According to \cite{Nelson:74}, {\em search goods} are products for which the consumer can obtain information about quality prior to purchase. Differently, {\em experience goods} require sampling or purchase to evaluate their quality.}
    \item 
    The second factor is the {\bf operationalization of perceived helpfulness}; in other words, its implementation. 
    \cite{Wu:17} analyzed \cite{Amazon} experience products and found that, considering the ratio between the number of positive votes and the total number of votes, review valence (intended as rating) positively influences helpfulness. However, the opposite result is obtained if helpfulness is computed as the count of votes received by a review. Moreover, \cite{Hong-etal:17} discovered that, while review length is a significant determinant of helpfulness, regardless of its operationalization, it has stronger effect when it is measured as the count of votes. More generally, \cite{Hong-etal:17} found that review length, review age and reviewer expertise positively influence perceived helpfulness while readability and rating are insignificant determinants, regardless of the applied helpfulness measure.
\end{itemize}
In our experiments we focus on accommodation and food services, all of which are classified as search products. Therefore, our analysis is not particularly affected by the effect of product moderation. Moreover, we operationalize perceived helpfulness as the count of votes because we use review datasets which include positive feedback about reviews. 

\subsection{Summary of our work}
We focus on review-related determinants and we leave apart reviewer-related properties (expertise, reputation, productivity, anonymity, trustworthiness, etc. \citep{Malik-Hussain:18,Filieri-etal:18,Siering-etal:18,Davis-Agrawal:18}) and context (review age, visibility, etc., \citep{Hu-Chen:16,Hu-etal:17}). We exclude these aspects in order to restrict the number of factors to be analyzed. 
We also exclude the analysis of the hedonic value of reviews \citep{Ham-etal:19} because we focus on decision-making-related aspects.

While some works have studied the deviation between the rating of a review and the mean rating of the item involved, our work introduces the deviations with respect to length, polarity and coherence, by user and by item, in order to understand whether this is helpful information to item evaluation. Moreover, we provide a prediction model that we validate by means of correlation analysis using observed perceived helpfulness, and by applying the model to a collaborative recommendation algorithm.

\section{Research Methodology}
\label{sec:method}
In order to focus on a set of largely-recognized helpfulness determinants, we select length, Unigram, rating, polarity and coherence as basic factors to be investigated and we study the deviations in the values of these factors from average, user-based or item-based. Specifically, we abstract from domain-dependent semantic concepts, which lack generalizability, and we only exploit Unigram as a lightweight measure for the assessment of the amount of content provided by reviews. We also exclude review age because it is a partial indicator, as the datasets we use provide no information about the visibility of reviews \citep{Hu-Chen:16}.

We analyse dependencies between factors and perceived helpfulness by exploiting regression models to understand the influence of determinants. Other works employ neural networks in learning data to use it for prediction models. See \cite{Fan-etal:19} and \cite{Malik-Hussain:17}. However, those approaches fail to shed light on the impact of individual features. In other words, they discover which combination of factors achieves the best results but they cannot reveal the influence of individual determinants on review helpfulness. Moreover, comparing a regression-based approach \citep{Yang-etal:16} with an advanced neural one \citep{Chen-etal:19} on the same dataset shows that the regression model performs almost as well as the neural one. While this might not be true in general, we prefer regression because of the transparency of its results.

Below, we introduce notation used in the following sections:
\begin{itemize} 
\item 
${\cal I} = \{i_1, \dots, i_m\}$ is the set of items (products or services);
\item  
${\cal U} = \{u_1, \dots, u_n\}$ is the set of users  - users can post reviews about items and vote the helpfulness of the reviews written by the other people;
\item 
${\cal R} = \{r_1, \dots, r_k\}$ is the set of reviews. We assume that each comment is associated with a rating of the reviewed item.
\end{itemize}

\begin{table}[]
\centering
\caption{Descriptive statistics of variables.}
\label{tab:descriptive-statistics}
\resizebox{\textwidth}{!}{%
{\def\arraystretch{1.3}
\begin{tabular}{@{}lllllllllllll@{}}
\toprule
\multicolumn{1}{l}{}                                                           & \multicolumn{6}{c}{Yelp-Hotel}                                          & \multicolumn{6}{c}{Yelp-Food}                           \\                         \cmidrule(lr){2-7} \cmidrule(l){8-13} 
                                                            
                                                                                & Count                & Min     & Max      & Mean     & STD      & Median & Count & Min    & Max      & Mean     & STD      & Median \\ \midrule
Number of reviews (+ ratings)                                                               & 10081                &         &          &          &          &        & 65120 &        &          &          &          &        \\
Number of users                                                                 & 654                  &         &          &          &          &        & 3105  &        &          &          &          &        \\
Number of items                                                                 & 1081                 &         &          &          &          &        & 2150  &        &          &          &          &        \\
Number of reviews x user                                                        &                      & 10      & 106      & 15.4144  & 9.1463   & 13     &       & 10     & 320      & 20.9726  & 18.2355  & 15     \\

Number of reviews x item                                                        &                      & 1      & 222      & 9.3256  & 27.1735   & 2     &       & 1     & 485      & 30.2884  & 49.4937  & 12     \\

\begin{tabular}[l]{@{}l@{}}\hspace{-1.5mm}
\vspace{-4mm}Number of helpfulness
\\
votes x review \end{tabular}

 &                      & 0       & 559      & 7.3118   & 18.1625  & 3      &       & 0      & 227      & 4.5535   & 9.1838   & 2      \\
Rating values                                                                   &                      & 1       & 5        & 3.5604   & 1.0661   & 4      &       & 1      & 5        & 3.7904   & 1.1519   & 4      \\
Review length                                                            &                      & 4       & 1005     & 175.5369 & 145.6903 & 134    &       & 1      & 1011     & 137.2749 & 117.2275 & 104    \\
Review polarity                                                          &                      & 1.1047  & 4.9310   & 3.9499   & 0.5796   & 4.1446 &       & 1.0989 & 4.9681   & 4.0086   & 0.5583   & 4.1774 \\
Rating-polarity coherence                                                       &                      & 1.6385  & 5.0000   & 4.2291   & 0.6276   & 4.3157 &       & 1.1721 & 5        & 4.2388   & 0.6135   & 4.3364 \\
STD of rating values x user                                                     &                      & 0       & 1.9315   & 0.9614   &          &        &       & 0      & 2.0248   & 1.0743   &          &        \\
STD of rating values x item                                                     & \multicolumn{1}{l}{} & 0       & 2.8284   & 0.7928   &          &        &       & 0      & 2.8284   & 1.0827   &          &        \\
STD of review polarity x user                                                   & \multicolumn{1}{l}{} & 0.0485  & 1.2782   & 0.4741   &          &        &       & 0.0444 & 1.3151   & 0.4977   &          &        \\
STD of review polarity x item                                                   & \multicolumn{1}{l}{} & 0.0002  & 1.6106   & 0.3931   &          &        &       & 0.0006 & 1.7474   & 0.5441   &          &        \\
STD of review length x user                                                     & \multicolumn{1}{l}{} & 12.0504 & 303.2880 & 94.3964  &          &        &       & 3.4667 & 325.8014 & 72.2702  &          &        \\
STD of review length x item                                                     & \multicolumn{1}{l}{} & 0.7071  & 536.6940 & 117.1095 &          &        &       & 0      & 627.9108 & 99.2652  &          &        \\  \bottomrule
\end{tabular}%
}}
\end{table}

\subsection{Data}
\label{sec:data-collection}
For our analysis we use two subsets of the \citep{Yelp-dataset} dataset: 
\begin{itemize}
    \item 
    YELP-Hotel stores reviews about accommodation services;
    \item 
    YELP-Food stores reviews about food services in the city of Phoenix.
\end{itemize}
In \citep{Yelp-dataset}, each item (business) is associated with a list of tags representing the categories to which it belongs. We obtained these datasets from the main one by applying two filters. First, we filtered items by tag and we removed the items which had no associated review+rating. Then, we removed the information about the users who provided less than 10 reviews. This is important to support the analysis of deviations in individual user behavior.\footnote{The full list of Yelp categories is available at  \url{https://www.yelp.com/developers/documentation/v3/category_list}. Appendix A reports the categories we used to produce the two datasets.}

In the datasets, each business is associated with the rating scores and free text reviews provided by Yelp users. Item ratings take values in a [1,5] Likert scale where 1 is the worst value and 5 is the best one. Moreover each review is associated with the feedback it receives from its own readers (for example, ``useful'' votes). Yelp only supports the expression of positive feedback.
Table \ref{tab:descriptive-statistics} provides some descriptive statistics about the two datasets:
\begin{itemize}
    \item 
    The higher portion of the table reports general statistics about the dataset (``Number of reviews (+ ratings)'', \dots, ``Number of helpfulness votes $\times$ review''). Looking at line ``Number of reviews $\times$ user'', which shows how many comments have been provided by individual users, we notice that this distribution has a long tail. Few users wrote many reviews; the other people provided very few ones. The distributions of the number of reviews $\times$  item and that of helpfulness votes $\times$ review are similar.
    \item 
    The second portion of the table (``Rating values'', \dots, ``Rating-Polarity coherence'') summarizes the distribution of rating scores on items, and statistics regarding review length and polarity. We compute the polarity of each comment as follows: 
    \begin{itemize}
        \item 
        First, we retrieve the polarity values generated by TextBlob \citep{TextBlob} and VADER \citep{Hutto-Gilbert:14};
        \item 
        Then, we compute the mean value among the two and we convert it in the [1, 5] interval.
    \end{itemize}
    In this way, the final polarity value can be compared with the rating associated to the review for the evaluation of the Rating-polarity coherence. We notice that reviews are fairly consistent, with a value of 4.229 in the [1, 5] interval.
    \item 
    The third portion of the table (``STD of rating values $\times$ user'', \dots, ``STD of review length $\times$ item'') provides information about the standard deviation of rating scores, polarity and length across users or items. Rating and reviewing behavior is not uniform. Specifically, we observe differences in the expression of ratings (mean STD=0.96 in [1, 5]), polarity (mean STD=0.474) and length (mean STD = 94 words). Analogous considerations can be made looking at reviews from the viewpoint of items, even though, in that case, the standard deviation is a bit lower.
\end{itemize}
The observations concerning standard deviations in reviews written by the same users, or concerning the same items, highlight the relevance of studying these aspects. In the following we describe the dependent and independent variables we defined and the analysis method we applied.

\subsection{Research variables}
\subsubsection{Dependent variable}
Given a review $r \in {\cal R}$ concerning an item $i \in {\cal I}$, the only dependent variable we consider is the {\bf perceived helpfulness} of $r$, which we operationalize in terms of counting votes, normalized in the [0, 1] interval:
\begin{equation}
    PerceivedHelpfulness_r = f(|Votes_r|)
    \label{eq:helpfulness}
\end{equation}
In Equation \ref{eq:helpfulness}, $Votes_r$ is the total number of votes received by $r$. This is the sum of ``useful'', ``funny'' and ``cool'' votes given to $r$. Moreover $|.|$ is set cardinality. Function $f()$ normalizes its argument in the [0, 1] interval:
\begin{equation}
    f(x) = \frac{log(x + 1)}{1 + log(x + 1)}
    \label{eq:normalization}
\end{equation}
We adopt this operationalization because, as previously specified, Yelp only supports positive feedback. 

\begin{table*}[!t]
\centering 
\caption{Independent variables and their operationalization: $r \in {\cal R}$ denotes a review about an item $i \in {\cal I}$ and $u \in {\cal U}$ is the author of $r$. $Words_r$ is the set of words included in $r$. $Revs_u$ is the set of reviews written by $u$. $Revs_i$ is the set of reviews about $i$ and $polarity_r$ is the polarity of $r$ computed as explained in Section \ref{sec:data-collection}. All the variables are normalized in [0,1] using formula $f()$ of Equation \ref{eq:normalization}.}
\resizebox{0.98\textwidth}{!}{%
{\def\arraystretch{2}
\begin{tabular}{lllll}

\toprule
\vspace{4mm}
Variable & Operationalization & Notes\\
\midrule

$RAT_r$ & 
$RAT_r = f(rating_r) $ & $rating_r$ is the rating included in $r$\\

$LEN_r$ & 
    $LEN_r = f(|Words_r|)$ & $|.|$ denotes set cardinality \\
    
$UGR_r$ &
\multicolumn{1}{l}{
 \renewcommand{\arraystretch}{0.7}
 \begin{tabular}[l]{@{}c@{}}
  $UGR_r = f(TF\_IDF)$ 
 \end{tabular}} &
 
\multicolumn{1}{l}{
 \renewcommand{\arraystretch}{0.7}
 \begin{tabular}[l]{@{}c@{}}
  $TF\_IDF_r = \frac{\sum\limits_{w\in Words_r} TF\_IDF_w}{|Words_r|}$ 
 \end{tabular}} \\

$POL_r$ & $f(polarity_r)$ &  \\

$COH_r$ & 
$f(1 - ||RAT_r - POL_r||)$ &
$||.||$ denotes absolute value \\

$\Delta{LEN_{ru}}$ & 
$f(||LEN_r - \frac{\sum\limits_{x\in Revs_u} LEN_x}{|Revs_u|}||)$ &  \\

$\Delta{LEN_{ri}}$ & 
$f(||LEN_r - \frac{\sum\limits_{x\in Revs_i} LEN_x}{|Revs_i|}||)$ &  \\

$\Delta{RAT_{ru}}$ & 
$f(||RAT_r - \frac{\sum\limits_{x\in Revs_u} RAT_x}{|Revs_u|}||)$ & \\

$\Delta{RAT_{ri}}$ & 
$f(||RAT_r - \frac{\sum\limits_{x\in Revs_i} RAT_x}{|Revs_i|}||)$ & \\

$\Delta{POL_{ru}}$ & 
$f(||POL_r - \frac{\sum\limits_{x\in Revs_u} POL_x}{|Revs_u|}||)$ & \\

$\Delta{POL_{ri}}$ & 
$f(||POL_r - \frac{\sum\limits_{x\in Revs_i} POL_x}{|Revs_i|}||)$ & \\

 \bottomrule
\end{tabular}
}
}
\label{tab:determinants}
\end{table*}

\subsubsection{Independent variables} 
Given a review $r \in {\cal R}$ about an item $i \in {\cal I}$, we consider the following independent variables. Table \ref{tab:determinants} shows the operationalizations of these variables, all of which are normalized in [0, 1] using Equation \ref{eq:normalization}:
\begin{itemize}
    \item 
    {\bf $RAT_r$:} normalized rating of $i$ in $r$. This is the normalized score value that $r$'s author attributed to item $i$;
    \item 
    {\bf $LEN_r$:} normalized number of words included in $r$;
    \item 
     $UGR_r$ \textit{(Unigram)}: normalized, mean TF/IDF value \citep{Robertson:04} of the lemmatized words included in $r$ after having removed stop words and very short words (composed of maximum 2 letters) from the text;
    \item 
    {\bf $POL_r$:} normalized polarity of the text of $r$, computed as explained in Section \ref{sec:data-collection}; 
    \item 
    {\bf $COH_r$:} normalized coherence between the polarity of $r$ and the rating of item $i$;
    \item 
    {\bf $\Delta{LEN_{ru}}$:} normalized absolute distance between the length of $r$ and the mean length of the reviews written by $u$;
    \item 
    {\bf $\Delta{LEN_{ri}}$:} normalized absolute distance between the length of $r$ and the mean length of the reviews about item $i$;
    \item 
    {\bf $\Delta{RAT_{ru}}$:} normalized absolute distance from the rating of $i$ in $r$ and the mean rating of items in the reviews written by $u$;
    \item 
    {\bf $\Delta{RAT_{ri}}$:} normalized absolute distance between the rating of $i$ in $r$ and the mean rating of $i$;
    \item 
    {\bf $\Delta{POL_{ru}}$:} normalized absolute distance between the polarity of $r$ and the mean polarity of the reviews written by $u$;
    \item 
    {\bf $\Delta{POL_{ri}}$:} normalized absolute distance between the polarity of $r$ and the mean polarity of the reviews about $i$.
\end{itemize}

\begin{table}[!t]
\centering
\caption{Pearson correlation coefficients between variables on Yelp-Hotel dataset.}
\label{tab:matrix-hotel}
\resizebox{\textwidth}{!}{%
\begin{tabular}{@{}llllllllllllll@{}}
\toprule
   &                     & 1       & 2       & 3       & 4       & 5       & 6       & 7       & 8      & 9      & 10      & 11      & 12 \\ \midrule
1  & $RAT_r$             & 1       &         &         &         &         &         &         &        &        &         &         &    \\
2  & $LEN_r$             & -0.0276 & 1       &         &         &         &         &         &        &        &         &         &    \\
3  & $UGR_r$             & 0.0109  & -0.9614 & 1       &         &         &         &         &        &        &         &         &    \\
4  & $POL_r$             & 0.5297  & 0.0525  & -0.0644 & 1       &         &         &         &        &        &         &         &    \\
5  & $COH_r$             & 0.6446  & -0.0436 & 0.0320  & 0.0856  & 1       &         &         &        &        &         &         &    \\
6  & $\Delta{LEN_{ru}}$ & -0.0487 & 0.1592  & -0.1340 & -0.0338 & -0.0385 & 1       &         &        &        &         &         &    \\
7  & $\Delta{LEN_{ri}}$ & -0.0018 & -0.0529 & 0.0761  & -0.0158 & -0.0046 & 0.0566  & 1       &        &        &         &         &    \\
8  & $\Delta{RAT_{ru}}$ & -0.3257 & 0.0197  & -0.0156 & -0.2151 & -0.4185 & 0.0494  & 0.0265  & 1      &        &         &         &    \\
9  & $\Delta{RAT_{ri}}$ & -0.2840 & -0.0272 & 0.0408  & -0.2113 & -0.3279 & 0.0162  & 0.3447  & 0.4061 & 1      &         &         &    \\
10 & $\Delta{POL_{ru}}$ & -0.2994 & -0.1782 & 0.1791  & -0.6683 & -0.0078 & -0.0015 & 0.0449  & 0.2616 & 0.1942 & 1       &         &    \\
11 & $\Delta{POL_{ri}}$ & -0.3158 & -0.1406 & 0.1510  & -0.6649 & -0.0373 & -0.0034 & 0.2582  & 0.1456 & 0.3306 & 0.6383  & 1       &    \\
12 & $PerceivedHelpfulness_r$          & -0.0531 & 0.3619  & -0.3567 & -0.0133 & -0.0625 & 0.1140  & -0.0026 & 0.0707 & 0.0415 & -0.0696 & -0.0428 & 1  \\ \bottomrule
\end{tabular}%
}
\end{table}

\begin{table}[!t]
\centering
\caption{Pearson correlation coefficients between variables on Yelp-Food dataset.}
\label{tab:matrix-food}
\resizebox{\textwidth}{!}{%
\begin{tabular}{@{}llllllllllllll@{}}
\toprule
   &                     & 1       & 2       & 3       & 4       & 5       & 6       & 7       & 8      & 9      & 10      & 11      & 12 \\ \midrule
1  & $RAT_r$             & 1       &         &         &         &         &         &         &        &        &         &         &    \\
2  & $LEN_r$             & -0.1058 & 1       &         &         &         &         &         &        &        &         &         &    \\
3  & $UGR_r$             & 0.0728  & -0.9594 & 1       &         &         &         &         &        &        &         &         &    \\
4  & $POL_r$             & 0.5879  & 0.0368  & -0.0614 & 1       &         &         &         &        &        &         &         &    \\
5  & $COH_r$             & 0.6323  & -0.0891 & 0.0679  & 0.1385  & 1       &         &         &        &        &         &         &    \\
6  & $\Delta{LEN_{ru}}$ & -0.0616 & 0.1506  & -0.1425 & -0.0166 & -0.0659 & 1       &         &        &        &         &         &    \\
7  & $\Delta{LEN_{ri}}$ & 0.0117  & -0.0825 & 0.0890  & 0.0249  & -0.0162 & 0.1076  & 1       &        &        &         &         &    \\
8  & $\Delta{RAT_{ru}}$ & -0.4164 & 0.0206  & -0.0061 & -0.3053 & -0.5346 & 0.0490  & 0.0117  & 1      &        &         &         &    \\
9  & $\Delta{RAT_{ri}}$ & -0.3944 & -0.0056 & 0.0200  & -0.2846 & -0.5443 & 0.0180  & 0.0553  & 0.5992 & 1      &         &         &    \\
10 & $\Delta{POL_{ru}}$ & -0.3517 & -0.2063 & 0.2173  & -0.6771 & -0.0428 & -0.0205 & 0.0050  & 0.3298 & 0.2239 & 1       &         &    \\
11 & $\Delta{POL_{ri}}$ & -0.3571 & -0.2052 & 0.2218  & -0.6837 & -0.0313 & -0.0292 & 0.0243  & 0.2100 & 0.2915 & 0.7193  & 1       &    \\
12 & $PerceivedHelpfulness_r$         & -0.0720 & 0.3685  & -0.3702 & -0.0319 & -0.0719 & 0.1467  & -0.0026 & 0.0510 & 0.0319 & -0.0663 & -0.0604 & 1  \\ \bottomrule
\end{tabular}%
}
\end{table}

\subsubsection{Correlation analysis}
Table \ref{tab:matrix-hotel} shows the Pearson correlation coefficient between each couple of independent variables and between the variables and the observed helpfulness ($PerceivedHelpfulness_r$) in the Yelp-Hotel dataset. Table \ref{tab:matrix-food} provides the same type of information for the Yelp-Food dataset. 
In the following, we jointly discuss the two sets of results because we observe similar correlation results.

Review polarity ($POL_r$) is highly correlated with the associated rating $RAT_r$ ($r_{Pearson}=0.5297$ in Yelp-Hotel, $r_{Pearson}=0.5879$ in Yelp-Food). This finding is consistent with the good ``Rating-polarity coherence'' observed in Table \ref{tab:descriptive-statistics} and suggests that the rating is mostly in line with the valence of review text.
Moreover $\Delta{RAT_{ri}}$ is highly correlated with $\Delta{RAT_{ru}}$ ($r_{Pearson}=0.4061$ in Yelp-Hotel, $r_{Pearson}=0.6992$ in Yelp-Food). Thus, the difference between the rating of an item $i$ and its average rating, and the difference between the rating of $i$ and the mean ratings provided by the same user, are similar.
The tables also show that there is a correlation between $\Delta{POL_{ri}}$ and $\Delta{POL_{ru}}$, denoting that a similar behavior is observed for review valence.
Finally, $PerceivedHelpfulness_r$ correlates well with review length ($LEN_r$).

As described in Section \ref{sec:empiricalModel}, in order to understand how variables interact with each other, and whether we can ignore some of them in perceived helpfulness estimation, we perform a regression analysis on the models that are based on these variables.

\subsection{Empirical model}
\label{sec:empiricalModel}
We consider two baseline helpfulness estimation models that include traditional determinants, and our proposed one:
\begin{itemize}
    \item 
    M1: $\beta_0$ + $\beta_1 RAT_r + \beta_2 LEN_r + \beta_3 UGR_r$;
    \item 
    M2: $\beta_0$ + $\beta_1 RAT_r + \beta_2 LEN_r + \beta_3 UGR_r + \beta_4 POL_r$;
    \item 
    M3: $\beta_0$ + $\beta_1 RAT_r + \beta_2 LEN_r + \beta_3 UGR_r + \beta_4 POL_r + 
    \beta_5 COH_r + 
    \beta_6 \Delta{LEN_{ru}} + \beta_7 \Delta{LEN_{ri}} + 
    \beta_8 \Delta{RAT_{ru}} + \beta_9 \Delta{RAT_{ri}} +
    \beta_{10} \Delta{POL_{ru}} + \beta_{11} \Delta{POL_{ri}}$.
\end{itemize}
We learn two versions of each model M$j$ ($j \in \{1, 2, 3\}$): 
\begin{enumerate} 
\item 
M$j_{L}$ is obtained by means of linear regression. For this version, we use Linear Support Vector Regression implemented in the \texttt{scikit-learn} library \citep{Pedregosa-etal:11}, which supports the analysis of the positive or negative influence of factors on helpfulness perception.
\item 
M$j_{NL}$ is obtained by means of a regression algorithm which can identify non linear dependencies. For this version, we use Random Forest Regression implemented in the \texttt{scikit-learn} library \citep{Pedregosa-etal:11}.
\end{enumerate}
We evaluate the models by comparing the estimated helpfulness values they generate with the helpfulness observed in the datasets by means of Pearson and Spearman correlation analyses.

\begin{table}[!t]
\centering
\caption{Pearson and Spearman correlation values of the M1, M2 and M3 models learned on Yelp-Hotel and Yelp-Food using Linear Support Vector Regression and Random Forest Regression. The best results are in bold. Significance is encoded as (**) $p<0.01$.}
\begin{tabular}{@{}lllll@{}}
\toprule
            & \multicolumn{2}{c}{Yelp-Hotel}                                                                                                 & \multicolumn{2}{c}{Yelp-Food}                                                                                                  \\ \cmidrule(lr){2-3} \cmidrule(l){4-5} 
                  &      Pearson's $r$ 
            
         &    Spearman's $r$ &    Pearson's $r$  &   Spearman's $r$ \\ \midrule
$M1_{L}$    & 0.3459**                                                        & 0.3652**                                                         & 0.3827**                                                        & 0.4114**                                                        \\
$M2_{L}$    & 0.3460**                                                        & 0.3644**                                                         & 0.3849**                                                        & 0.4137**                                                         \\
$M3_{L}$  & 0.3635**                                                        & 0.3725**                                                         & 0.3982**                                                        & 0.4169**                                                         \\ \midrule
$M1_{NL}$   & 0.2952**                                                        & 0.3002**                                                         & 0.3185**                                                        & 0.3240**                                                         \\
$M2_{NL}$   & 0.3065**                                                        & 0.3106**                                                         & 0.3675**                                                        & 0.3729**                                                         \\
$M3_{NL}$ & \textbf{0.4071**}                                               & \textbf{0.4036**}                                                & \textbf{0.4418**}                                               & \textbf{0.4486**}                                                \\ \bottomrule
\end{tabular}
\label{tab:results-regression}
\end{table}

\section{Results}
\label{sec:results}

\subsection{Performance of the helpfulness estimation models}
Table \ref{tab:results-regression} show the results of Linear Support Vector Regression and Random Forest Regression applied to M1, M2 and M3 on Yelp-Hotel and Yelp-Food.
For each dataset we carried out the experiments by applying a 5-fold cross validation in order to avoid biases related to the way the dataset is split. 
We can notice that M3 outperforms M1 and M2 in terms of Pearson and Spearman correlation both when learned through linear (M$3_{L}$) and non linear (M$3_{NL}$) regression.
This means that, by adding the deviations of rating, polarity and length to the variables used in M$_1$ and M$_2$, we improve helpfulness prediction.
Overall, M3$_{NL}$ obtains the best results. In other words, the Random Forest Regressor is a better helpfulness predictor than Linear Support Vector Regression on the datasets we analyzed. 

\begin{table}[!t]
\centering
\caption{Analysis on the two datasets by means of Linear Support Vector Regression. For each model, the coefficients show the impact of variables on perceived helpfulness. Significance is encoded as (**) $p<0.01$ and (*) $p<0.05$.}
\label{tab:weights-svr}
\resizebox{\columnwidth}{!}{%
{\def\arraystretch{1.5}
\begin{tabular}{@{}lllllll@{}}
\toprule
  \multicolumn{1}{@{}l}{\multirow{2}{*}{Variable}}            & \multicolumn{3}{c}{Yelp-Hotel} & \multicolumn{3}{c}{Yelp-Food}  \\ \cmidrule(lr){2-4} \cmidrule(l){5-7}
                   \multicolumn{1}{l}{}  & $M1_{L}$ & $M2_{L}$ & $M3_{L}$ & $M1_{L}$ & $M2_{L}$ & $M3_{L}$ \\ \midrule
$RAT_r$             & -0.1905$^{**}$  & -0.1560$^{**}$  &          & -0.1662$^{**}$  & -0.0741$^{**}$  & 0.0963$^{**}$   \\
$LEN_r$             & 1.5740$^{**}$   & 1.5836$^{**}$   &          & 0.7060$^{**}$   & 0.7016$^{**}$   & 0.6275$^{**}$   \\
$UGR_r$             & -0.8825  & -0.8820  & -4.1487$^{**}$  & -2.5234$^{**}$  & -2.5789$^{**}$  & -2.4575$^{**}$  \\
$POL_r$             &          & -0.1330  & -0.5983$^{**}$  &          & -0.3712$^{**}$  & -0.7557$^{**}$  \\
$COH_r$             &          &          & -0.1008  &          &          & -0.2534$^{**}$  \\
$\Delta{LEN_{ru}}$ &          &          & 0.1395$^{**}$   &          &          & 0.1814$^{**}$   \\
$\Delta{LEN_{ri}}$ &          &          & 0.0222   &          &          &          \\
$\Delta{RAT_{ru}}$ &          &          & 0.1041$^{**}$   &          &          & 0.0520$^{**}$   \\
$\Delta{RAT_{ri}}$ &          &          & 0.0571$^{*}$   &          &          &          \\
$\Delta{POL_{ru}}$ &          &          & -0.1500$^{**}$  &          &          & -0.1231$^{**}$  \\
$\Delta{POL_{ri}}$ &          &          & -0.0455  &          &          &          \\ \bottomrule
\end{tabular}
}}
\end{table}

\subsection{Impact of independent variables on review helpfulness}
\subsubsection{Linear SVR regression results}
Table \ref{tab:weights-svr} shows the weights assigned to the variables by the Linear Support Vector Regression on Yelp-Hotel and Yelp-Food. Most coefficients are statistically significant. 
First, we analyze the perceived helpfulness determinants identified in the previous research. Then we study the ones we propose.

On both datasets, $UGR_r$ negatively influences perceived helpfulness in all the models and the same happens for $POL_r$ in M2$_L$ and M3$_L$.
Furthermore $RAT_r$ has a negative influence, with the exception of M3$_L$ where this factor is not used, or it has a slightly positive effect.
Differently, $LEN_r$ has a positive impact on helpfulness in all the models, except for M3$_L$ that does not use it on Yelp-Hotel. These results are fairly consistent with the previous research results, some of which have investigated the importance  of these factors, but not the positive nor negative direction of influence.

As far as models M2$_L$ and M3$_L$ are concerned, we observe that $COH_r$ is used on both datasets with a negative weight. While this is apparently counterintuitive, it must be noted that the consistency between review sentiment and rating does not mean that the review is helpful. In fact, in the recommender systems research, this type of information is used to assess reviewers' reliability, which is different from review helpfulness \citep{Shen-etal:2019}. 

The results concerning  the deviations on length, rating and polarity are not completely aligned but they are fairly consistent: 
\begin{itemize}
    \item 
    On Yelp-Hotel the regression model uses all these factors for prediction. While the deviations in polarity have negative impact on perceived helpfulness, the deviations in length and rating positively influence this variable. We also observe that the user-based deviations have stronger influence than item-based ones, whose impact is low;
    \item 
    On Yelp-Food the impact of user-based length and rating deviations are consistent with those of the other dataset, while item-based deviations are not used.
\end{itemize}

\begin{table}[!t]
\centering
\caption{Analysis on the two datasets using Random Forest Regression. For each model, the values show the importance of the corresponding variables.}
\label{tab:weights-rf}
\begin{tabular}{@{}lllllll@{}}
\toprule
\multicolumn{1}{@{}l}{\multirow{2}{*}{Variable}}          & \multicolumn{3}{c}{Yelp-Hotel}    & \multicolumn{3}{c}{Yelp-Food}     \\  \cmidrule(lr){2-4} \cmidrule(l){5-7}
                 \multicolumn{1}{l}{}      & M1$_{NL}$ & M2$_{NL}$ & M3$_{NL}$ & M1$_{NL}$ & M2$_{NL}$ & M3$_{NL}$ \\ \midrule
$RAT_r$             & 0.0352    & 0.0637    & 0.0346    & 0.0187    & 0.0435    & 0.0208    \\
$LEN_r$             & 0.4129    & 0.3023    & 0.1949    & 0.3206    & 0.2571    & 0.1834    \\
$UGR_r$             & 0.5520    & 0.3574    & 0.2257    & 0.6608    & 0.4130    & 0.2400    \\
$POL_r$             &           & 0.2766    &           &           & 0.2864    &           \\
$COH_r$             &           &           &           &           &           & 0.1313    \\
$\Delta{LEN_{ru}}$ &           &           & 0.1559    &           &           & 0.1602    \\
$\Delta{LEN_{ri}}$ &           &           &           &           &           &           \\
$\Delta{RAT_{ru}}$ &           &           & 0.1378    &           &           & 0.1299    \\
$\Delta{RAT_{ri}}$ &           &           & 0.1202    &           &           &           \\
$\Delta{POL_{ru}}$ &           &           &           &           &           & 0.1343    \\
$\Delta{POL_{ri}}$ &           &           & 0.1309    &           &           &           \\ \bottomrule
\end{tabular}

\end{table}

\subsubsection{Random Forest Regression results}
Table \ref{tab:weights-rf} shows the importance of the independent variables using Random Forest Regression.
$UGR_r$, which represents the content of the reviews, strongly affects helpfulness prediction in all the models and datasets. It is the most influential determinant among the traditional ones (rating, length and Unigram). 
$LEN_r$ is quite influential as well: it has the second highest importance in all the models. $RAT_r$ is used everywhere but with minor importance.

Interestingly, $POL_r$ is rather influent in M2$_{NL}$ on both datasets. However, it disappears in M3$_{NL}$, which exploits user-based and/or item-based deviations. This suggests that these factors are more predictive than polarity. Moreover, $COH_r$ is only used on Yelp-Food.

In M3$_{NL}$ the Random Forest Regressor uses different deviation variables for the prediction:
\begin{itemize}
    \item 
    $\Delta{LEN_{ru}}$ and $\Delta{RAT_{ru}}$ are used in both datasets;
    \item 
    $\Delta{POL_{ru}}$ is only recognized as influential in Yelp-Food;
    \item 
    $\Delta{RAT_{ri}}$ is only used in Yelp-Hotel;
    \item 
    $\Delta{LEN_{ri}}$ is ignored in both datasets;
    \item 
    $\Delta{POL_{ru}}$ is only used in Yelp-Food;
    \item 
    $\Delta{POL_{ri}}$ is only used in Yelp-Hotel.
\end{itemize}
This shows that, while some variables, such as rating, length and Unigram are influential, the impact of coherence and item-based ratings depends on product type. In particular, for Yelp-Food user-based deviations are meaningful while item-based ones are not. In contrast, the impact of user-based deviations, especially concerning review length and rating, is consistent across datasets.

Overall, we notice that variables representing deviations are less important than ratings, length and Unigram. However they clearly help improving perceived helpfulness prediction, as shown in Table \ref{tab:results-regression}.

\section{Helpfulness-aware, personalized item recommendation}
\label{sec:recommender}
As described by \cite{Ricci-etal:11}, recommender systems leverage data about users' past rating behavior to personalize the suggestion of items. However, they uniformly exploit this type of information to predict ratings, without considering its quality. We point out that the usage of poor data might decrease recommendation performance because item evaluation would be based on unreliable information. With this perspective, we can further check the efficacy of our helpful estimation model by comparing a state of the art recommender system with an algorithm that tunes the impact of ratings in item evaluation on the basis of the predicted helpfulness of reviews.
We develop our SVD$_{Helpfulness}$ recommender system by modifying SVD \citep{Koren-Bell:11}, which is a largely-used collaborative recommender system based on Matrix Factorization. See Sections \ref{sec:MF} and \ref{sec:MFHelpfulness}. Then, we compare the accuracy, error minimization and ranking capabilities of SVD$_{Helpfulness}$ with those of SVD++ \citep{Koren:08}, which is a well-established baseline to evaluate recommender systems. See Section \ref{sec:validation}.   
For the development of SVD$_{Helpfulness}$ we predict the helpfulness of reviews by means of a hybrid model. Given a review $r \in {\cal R}$, $PredictedHelpfulness_r$ is computed as follows: 
\begin{equation}\resizebox{0.9\hsize}{!}{$
     PredictedHelpfulness_r =
    \begin{cases}
        PerceivedHelpfulness_r & \quad \text{if } |Votes_r|>0 \\
        \text{value estimated by M3}_{NL} & \quad \text{otherwise}
    \end{cases}$}
    \label{eq:predictedHelpfulness}
\end{equation}
If the ground-truth helpfulness is available we use it. Otherwise, 
\linebreak
$PredictedHelpfulness_r$ is the value obtained by applying M3$_{NL}$, which is the best performing model according to the analysis of Section \ref{sec:results}. Notice that we tested other combinations of the two measures, including their average, but the final performance of the recommender was lower. Therefore we selected this approach for our experiments. Overall, we use the helpfulness values estimated by M3$_{NL}$ in 21\% of Yelp-Hotel
reviews, and in 30\% of Yelp-Food reviews.

\subsection{Collaborative Filtering with Matrix Factorization}
\label{sec:MF}
The recommender systems based on Matrix Factorization assume that a few latent patterns influence rating behavior. These systems perform a low-rank matrix factorization on the users-items rating matrix, which stores the evaluations of items provided by users \citep{Koren-Bell:11}. We assume that there are $n$ users and $m$ items and we adopt the following notation:
\begin{itemize} 
\item 
{\bf R} $\in {\rm I\!R}^{n \times m}$ is the users-items rating matrix;
\item 
${\bf R}_{xy}$ is the rating given by user $u_x \in {\cal U}$ to item $i_y \in {\cal I}$, if any:
\begin{itemize}
    \item 
    ${\cal O} = \{<u_x, i_y> |~ {\bf R}_{xy} \neq 0 \}$ is the set of observed ratings. This set includes all the $<u_x, i_y>$ pairs such that, as reported in the dataset, user $u_x$ has given a rating in [1, 5] to item $i_y$.
    \item
    ${\cal T} = \{<u_x, i_y> |~ {\bf R}_{xy} = 0 \}$ is the set of unknown ratings. 
\end{itemize}
\end{itemize}
We assume that there are $K$ latent factors; then:
\begin{itemize}
    \item 
    {\bf u}$_x \in {\rm I\!R}^K$ denotes the user preference vector of user $u_x$ and ${\bf U} = [{\bf u}_1, \dots, {\bf u}_n] \in {\rm I\!R}^{K \times n}$ stores the preference vectors of all the users;
    \item 
    {\bf i}$_y \in {\rm I\!R}^K$ denotes the item characteristic vector of $i_y$ and ${\bf I} = [{\bf i}_1, \dots, {\bf i}_m] \in {\rm I\!R}^{K \times m}$ stores the item characteristic vectors of all the items.
\end{itemize}

In order to learn these vectors, the recommender system solves the following optimization problem:
\begin{equation}
    \min\limits_{{\bf U}, {\bf I}} \sum\limits_{<u_x, i_y> \in {\cal O}} 
     ({\bf R}_{xy} - {\bf u}_x^T {\bf i}_y)^2 +
     \lambda(||{\bf U}||^2_ F + ||{\bf I}||^2_F)
    \label{eq:MatrixFactorization}
\end{equation}
Equation \ref{eq:MatrixFactorization} is aimed at finding a setting of ${\bf U}$ and ${\bf I}$ that minimizes the distance between the observed ratings (${\cal O}$) and the estimated ones (obtained as the product of transposed user preference vectors and item characteristic ones).
In the equation,
$||.||_F$ denotes the Frobenius Norm and $||{\bf U}||^2_ F + ||{\bf I}||^2_F$ are the regularization terms to avoid over-fitting.
Moreover $\lambda>0$ controls the impact of {\bf U} and {\bf I} on regularization. The smaller is  $\lambda>0$, the minor is the influence of these vectors.

\subsection{Steering Matrix Factorization by means of review helpfulness}
\label{sec:MFHelpfulness}
In order to steer Matrix Factorization by means of review helpfulness, we introduce a weighting factor that tunes the impact of ratings depending on the predicted helpfulness of the associated reviews. The idea is that the ratings associated to highly helpful reviews should influence Matrix Factorization more than the other ones. The optimization problem is thus:
\begin{equation}
    \min\limits_{{\bf U}, {\bf I}} \sum\limits_{<u_x, i_y> \in {\cal O}} 
    w_{xy} ({\bf R}_{xy} - {\bf u}_x^T {\bf i}_y)^2 +
     \lambda(||{\bf U}||^2_ F + ||{\bf I}||^2_F)
    \label{eq:MatrixFactorization2}
\end{equation}
where $w_{xy}$ is the predicted helpfulness of the review $r$ associated to rating ${\bf R}_{xy}$, corresponding to $PredictedHelpfulness_r$ in Equation \ref{eq:predictedHelpfulness}.

\subsection{Validation}
\label{sec:validation}
In this section we present the evaluation results of SVD$_{Helpfulness}$ compared with SVD++, using standard performance metrics for recommender systems \citep{Jannach-etal:16}.
\begin{table}[!t]
\centering
\caption{Recommendation performance of SVD$_{Helpfulness}$ and SVD++ on ``Yelp-Hotel'' and ``Yelp-Food'' datasets. Stars indicate significant differences according to a Wilcoxon signed-rank test between the two algorithms ((*)  $p<$0.05). The percentages in brackets denote the relative difference between the values obtained by the two algorithms.}
\label{tab:rec-results}
\begin{tabular}{@{}lllll@{}}
\toprule
              & \multicolumn{2}{c}{Yelp-Hotel} & \multicolumn{2}{c}{Yelp-Food} \\\cmidrule(lr){2-3} \cmidrule(l){4-5}
              & SVD$_{Helpfulness}$     & SVD++      & SVD$_{Helpfulness}$     & SVD++     \\ \midrule
Precision     & 0.7971* (+4.66\%)         & 0.7616     & 0.7804* (+2.46\%)        & 0.7617    \\
Recall        & 0.7458* (+3.03\%)         & 0.7239     & 0.7956* (+5.55\%)         & 0.7538    \\
F1            & 0.7706* (+3.81\%)        & 0.7423     & 0.7879* (+3.99\%)         & 0.7577    \\
MAP           & 0.7159* (+3.83\%)        & 0.6895     & 0.7385*  (+6.60\%)        & 0.6928    \\
MRR           & 0.6313\phantom{*} (+1.61\%)          & 0.6213     & 0.7678*  (+1.59\%)        & 0.7558    \\
NDCG          & 0.9782*  (+0.56\%)        & 0.9728     & 0.9666* (+0.62\%)         & 0.9606    \\
RMSE          & 0.9279* (-6.13\%)         & 0.9885     & 1.0655*  (-7.39\%)        & 1.1505    \\
MAE           & 0.7206* (-0.73\%)         & 0.7726     & 0.8314* (-7.47\%)         & 0.8985    \\
\bottomrule
\end{tabular}%
\end{table}
We focus on ranking capability (MAP, MRR and NDCG), which is very important in the evaluation of recommender systems because it tells us how good an algorithm is at placing relevant items in the first positions of the suggestion list. Moreover we consider accuracy (Precision, Recall and F1) and error minimization (MAE and RMSE) metrics.
For the evaluation, we perform a 5-fold cross validation, having set the number of latent factors to 50 and the learning rate to 0.01, and we optimize the other parameters by taking the configuration that achieves the best MAP as optimal.

Table \ref{tab:rec-results} shows the results of the two algorithms on Yelp-Hotel and Yelp-Food and reports the relative differences in performance as percentages. Notice that, different from the other metrics, RMSE and MAE describe the error in rating estimation and thus have to be minimized. Therefore, the negative values associated to these measure denote an improvement in performance. On both datasets, SVD$_{Helpfulness}$ outperforms SVD++ in all the measures.

We can provide a more general view of performance by grouping measures according to high-level dimensions and by computing the mean values of the relative differences between algorithms. SVD$_{Helpfulness}$ compares to SVD++ as follows:
\begin{itemize}
    \item 
    The mean relative improvement of accuracy (Precision, Recall and F1) is equal to 3.83\% on YELP-Hotel and by 3.99\% on YELP-Food;
    \item
    The mean relative improvement of ranking capability (MAP, MRR and NDCG) is equal to 1.99\% on Yelp-Hotel and 2.94\% on Yelp-Food;
    \item
    Finally, SVD$_{Helpfulness}$ outperforms SVD++ in error minimization (RMSE and MAE) by 6.43\% on YELP-Hotel and 7.43\% on Yelp-Food.
\end{itemize}

\section{Discussion}
\label{sec:discussion}
We investigated the impact on perceived review helpfulness of deviations in length, polarity and rating with respect to the reviews written by the same user, or concerning the same item. The results of our study show that the deviations from typical user behavior influence perceived helpfulness and support the identification of high-quality ratings in collaborative recommendation.
By considering the results obtained using Random Forest Regression (which outperforms Linear Support Vector Regression in helpfulness estimation), we can answer our research questions as follows:
\begin{itemize}
    \item 
    {\em RQ1: Given a review r, does a deviation from the mean length, polarity and rating of the other reviews written by the same person provide useful information to assess the perceived helpfulness of r?}
    
    User-based deviations are useful to predict perceived review helpfulness. Specifically, the deviations concerning length and ratings influence helpfulness across both datasets that we considered, while the deviations in polarity have a more limited impact on it. These findings suggest that user-based deviations are important determinants to be considered within a review helpfulness estimation model. Particular attention should be given to the deviations in length and ratings which are, at the same time, stronger predictors and lighter measures to be computed than polarity. However, polarity should not be disregarded as a proxy of ratings in datasets that do not provide this type of information, such as the \cite{Airbnb} one available at \url{http://insideairbnb.com/get-the-data.html}.
    
    \item
    {\em RQ2: Given a review r, does a deviation from the mean length, polarity and rating of the other reviews on the same item provide useful information to assess the perceived helpfulness of r?}
    
    The situation of item-based deviations is more complex because the results across datasets are heterogeneous. Deviations in review length can be ignored because they have no impact on helpfulness estimation. Differently, deviations in rating and in polarity are only useful in the Yelp-Hotel dataset. Therefore, the value of these indicators must be assessed before applying them to analyze review helpfulness in a specific domain.
\end{itemize}

\subsection{Theoretical implications}
Overall, these results advance the state of the art in helpfulness estimation, which traditionally focused on local properties of reviews (\cite{Mudambi-Schuff:10,Yang-etal:15,Hong-etal:17,Siering-etal:18}), or on rating deviations with respect to the average evaluation of an item (\cite{Raghavan-etal:12,Fang-etal:16}). The novel aspect of our work is the analysis of consumer feedback on a broader user or item-related context, considering a larger set of determinants.

As discussed in Section \ref{sec:validation}, the validation carried out by integrating our helpfulness estimation model into collaborative recommendation shows that the SVD$_{Helpfulness}$ algorithm sensibly outperforms SVD++ (which uniformly applies ratings for item estimation) in accuracy, ranking capability and error minimization on both datasets we considered.
The superior performance of SVD$_{Helpfulness}$ is obviously relevant to the generation of good recommendation lists. Moreover, it has important implications regarding algorithmic experience \citep{Shin-etal:20} because it supports an explanation of suggestions based on high-quality feedback provided by the people who have previously experienced items. Specifically, a helpfulness-based exploitation of ratings and of the associated reviews can be the basis for the generation of explanations based on reliable user experience, as suggested by \cite{Ghose-Ipeirotis:11} and preliminarily investigated by \cite{Mauro-etal:20d}.

\subsection{Practical implications}
The findings of this study are expected to provide insights for retailers and platform developers regarding how to suggest helpful reviews to consumers, out of the plethora of available ones, in order to reduce information overload and to support decision-making. As discussed by \cite{Salehan-Kim:16}, older reviews tend to attract more readerships and many products have thousands of comments, most of which ``never receive any attention from consumers because they are at the end of a long list.'' Our idea is thus that of recommending reviews that provide useful content to support item selection and purchase decisions. Some online review platforms, such as \cite{TripAdvisor}, provide filters to explicitly select comments by facets (language, geographic location, etc.) but they overlook the usefulness of review content. Other platforms, such as \cite{Amazon} and \cite{Airbnb}, apply ranking strategies to promote the reviews which they consider as the most effective ones.  For instance, they show helpful positive and negative comments first. Moreover, some researchers have proposed heuristics aimed at re-ranking reviews in order to balance their visibility and to address the ``rich gets richer'' dilemma \citep{Wang-etal:20}. Our work advances the state of the art by identifying novel helpfulness determinants with the idea that ``out of the core'' messages can be relevant information sources for item selection. This criterion could be combined with the existing ones in order to promote valuable consumer feedback as soon as it is posted online, regardless of its own popularity. 

It is worth mentioning that we proposed a model aimed at identifying the most informative reviews to help consumers in decision-making. Another relevant perspective concerns retailers and service providers. In that case, consumer feedback can be analyzed to identify the positive and negative aspects of products and services observed by consumers, with the aim of highlighting aspects that can be improved or promoted. For instance, see \cite{Qi-etal:16}, \cite{Xu-Li:16}, \cite{Bilici-Saygin:17}, \cite{Prado-Moro:17} and \cite{Xu-etal:19}.
We are carrying out preliminary investigations in this direction to evaluate the helpfulness determinants we proposed in that context.

\section{Limitations and future work}
\label{sec:future}
Our work has limitations that we would like to address:
\begin{enumerate}
    \item 
    Helpfulness estimation also concerns the properties of reviewers like their expertise \citep{Siering-etal:18}, reputation \citep{Tang-etal:13b,Chua-Banerjee:15,Huang-etal:15,Gao-etal:17,Jiang-Diesner:16}, engagement and social influence \citep{Ngo-Ye-Sinha:14,Mohammadiani-etal:17}. Within this extensive scenario, we contribute to enhance the content-analysis perspective. However, we plan to extend our analysis to the other variables.
    \item 
    We tested our model on two datasets concerning food and accommodation services. Further experiments with different datasets are needed to assess the validity of the model in other domains, such as the sales of search and experience products.
    \item
    We describe the semantic features of reviews by focusing on TF/IDF to measure the relative importance of the words occurring in the reviews. In our future work we plan to integrate in our model an analysis of the role of emotions in order to enrich the type of information used for helpfulness assessment \citep{Martin-Pu:14,Yang-etal:15,Malik-Hussain:17,Gang-Taeho:19}.
\end{enumerate}
We also plan to extend our work regarding review-aware recommender systems \citep{Chen-etal:15,Rubio-etal:19}. Previously, \cite{Mauro-etal:19} and \cite{Ardissono-Mauro:20} investigated reviewer behavior with the aim of estimating reputation and they leveraged this information in a trust-based recommender system. Now, we plan to extend that work with the analysis of review helpfulness which, in turn, affects reviewers' trustworthiness.

Another interesting research path is the enhancement of data interpretation via information visualization, which we investigated in our recent research about information exploration \citep{Mauro-etal:19,Mauro-etal:20b}. This is relevant to reveal interesting behavior patterns related to the influence of context (for instance, travel context in \citep{Chang-etal:19b}) and cultural background \citep{Nakayama-Wan:19b} on the aspects and evaluations appearing in the reviews.

\section{Conclusions}
\label{sec:conclusions}
This paper presented a study on perceived review helpfulness aimed at advancing the state of the art in the identification of the reviews which provide useful information to consumers for decision-making. We proposed a novel perspective on review analysis by considering {\em deviations in rating, length and valence} with respect to the reviews written by the same user, or concerning the same item.

We studied this phenomenon by developing three models which leverage different sets of content features: from traditional ones (length, rating, Unigram and polarity) to the novel determinants we propose, including coherence between rating and polarity, and deviations in length, polarity and rating pivoted on users and on items. We learned these models through regression on two large datasets about accommodation and food services. Our experiments showed that the analysis of these factors enhances the estimation of review helpfulness by reaching higher correlation values with the helpfulness feedback observed in the datasets. Specifically, the experimental results show that user-based deviations, especially regarding review length and rating, influence perceived helpfulness, while item-based deviations are weaker predictors.
A further experiment in which we integrated helpfulness estimation into a collaborative recommender system has shown that this type of information enhances suggestion performance with respect to only using item ratings. In other words, the reviews estimated as helpful provide high-quality content for recommendation.

Overall these results are encouraging and suggest that our model is an effective tool to select relevant user feedback for decision-making.

\section{Acknowledgments}
This work was supported by the University of Torino (grant number: 
\linebreak 
ARDL\_RILO\_19\_01) which funded the conduct of the research and preparation of the article. We are grateful to Ms. Jeanne Marie Griffin for having proofread our paper.


\appendix
\section{}
\label{sec:appendix}
\begin{itemize}
\item 
The selection of businesses to define the Yelp-Hotel dataset is based on the following tags: Hotels,
Mountain Huts,
Residences,
Rest Stops,
Bed \& Breakfast,
Hostels,
Resorts. 
\item
The selection for Yelp-Food is based on the following tags:
American,
Argentine,
Asian Fusion,
Australian,
Austrian,
Bangladeshi,
Belgian,
Brasseries,
Brazilian,
British,
Cambodian,
Cantonese,
Catalan,
Chinese,
Conveyor Belt Sushi,
Cuban,
Czech,
Delis,
Empanadas,
Falafel,
Filipino,
Fish \& Chips,
French,
German,
Greek,
Hawaiian,
\linebreak
Himalayan/Nepalese,
Hot Pot,
Hungarian,
Iberian,
Indian,
Indonesian,
Irish,
Italian,
Japanese,
Japanese Curry,
Korean,
Latin American,
Lebanese,
Malaysian,
Mediterranean,
Mexican,
Middle Eastern,
Modern European,
Mongolian,
New Mexican Cuisine,
Noodles,
Pakistani,
Pan Asian,
Persian/Iranian,
Peruvian,
Piadina,
Pizza,
Poke,
Polish,
Polynesian,
Portuguese,
Ramen,
Russian,
Salad,
Scandinavian,
Scottish,
Seafood,
Shanghainese,
Sicilian,
Singaporean,
Soup,
Southern,
Spanish,
Sri Lankan,
Steakhouses,
Sushi Bars,
Syrian,
Tacos,
Tapas Bars,
Tapas/Small Plates,
Teppanyaki,
Tex-Mex,
Thai,
Turkish,
Ukrainian,
Vegan,
Vegetarian,
Vietnamese,
Wraps.
\end{itemize}


\clearpage










\end{document}